\documentclass[a4paper,11pt]{article}
\usepackage{pos}
\usepackage{wrapfig}

\title{Enabling distributed analysis for ALICE Run 3}
\author*[1]{Raluca Cruceru}

\affiliation{CERN,\\
  Espl. des Particules 1, 1211 Meyrin, Switzerland}

\note{On behalf of the ALICE Collaboration}

\emailAdd{raluca.cruceru@cern.ch}

\abstract{The ALICE Collaboration has just finished a major detector upgrade that increases the data-taking rate capability by two orders of magnitude and will allow to collect unprecedented data samples. For example, the analysis input for 1 month of Pb-Pb collisions amounts to about 5 PB. In order to enable analysis on such large data samples, the ALICE distributed infrastructure was revised and dedicated tools for Run 3 analysis were created. These are firstly the {$\mathrm{O^2}$} analysis framework that builds on a multi-process architecture exchanging a flat data format through shared memory implemented in C++. Secondly, the Hyperloop train system for distributed analysis on the Grid and on dedicated analysis facilities implemented in Java/Javascript/React. These systems have been commissioned with converted Run 2 data and with the recent LHC pilot beam and are ready for data analysis for the start of Run 3. This contribution discusses the requirements and the used concepts, providing details on the actual implementation. The status of the operation in particular with respect to the LHC pilot beam will also be discussed.}

\FullConference{%
  41st International Conference on High Energy physics - ICHEP2022\\
  6-13 July, 2022\\
  Bologna, Italy
}

\begin{document}
\maketitle

\section{Introduction}
The ALICE detector has undergone a number of significant upgrades that considerably increased the data-taking capability, as the main detector parts have been replaced and improved. Starting with the installation of a new Inner Tracking System and a new Fast-Interaction Trigger system, the Time Projection Chamber has also been upgraded in the process~\cite{ALICE:2012dtf}. Furthermore, a new Online-Offline data taking and processing infrastructure has been developed. 

Following these substantial developments, the ability to record data has increased by two orders of magnitude, allowing for 100 times more collisions recorded compared to Run 1 and Run 2. The resulting data throughput from the detector is estimated to be greater than 1 TB/s for Pb-Pb collisions. The reconstruction output, which is the analysis input, has an unprecedented size, too, for 1 month for Pb-Pb collisions it amounts to about 5 PB. 

In order to efficiently enable analysis on such large data samples, the ALICE distributed infrastructure was revised and dedicated tools for Run 3 analysis were created. The first one is the {$\mathrm{O^2}$} analysis framework~\cite{Alkin:2021mfo} implemented in C++, which builds on a multi-process architecture exchanging a flat data format through shared memory. The second tool is the Hyperloop train system for distributed analysis on the Grid and dedicated analysis facilities, which was implemented in Java/Javascript/React. These systems have been commissioned with converted Run 2 data and with the recent LHC pilot beam and are at present used for the analysis of Run 3 data.

\section{The \texorpdfstring{$\mathrm{O^2}$}{} Framework}

\begin{figure}[ht]
\centering
\vspace{5pt} 
\includegraphics[width=\textwidth, height=\textheight, keepaspectratio]{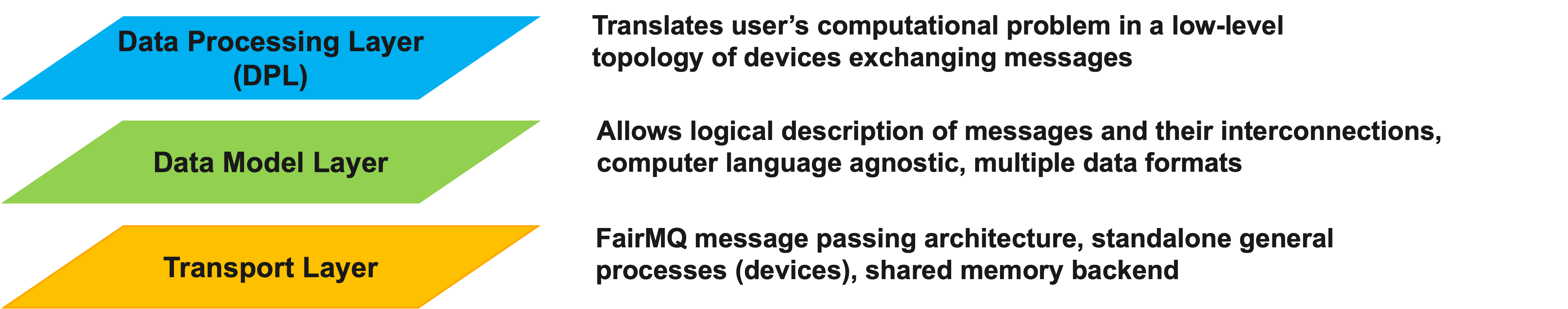}
  \caption{{$\mathrm{O^2}$} Framework Layers}
  \label{fig:O2layers}
\end{figure}

As dedicated Framework to tackle the challenges of LHC Run 3, {$\mathrm{O^2}$} is derived from the ALICE High-Level Trigger architecture – a message-passing multiprocess system used in Run 1 and 2, and allows for distributed and efficient processing of the new amount of data~\cite{Buncic:2015ari}. It involves two main processing stages: the synchronous processing (online), resulting in the CTF (Compressed Time Frame) - stored on the disk buffer, and the asynchronous stage (offline) that performs the reconstruction with the final calibration and produces the final AOD (Analysis Object Data). The CTF and AOD are then sent to permanent storage.
{$\mathrm{O^2}$} hides the underlying distributed framework, which allows analyzers to \mbox{focus on the physics. This is done through the three main components depicted in Figure~\ref{fig:O2layers}.}

\subsection{Data Model and table manipulation}

\begin{figure}[ht]
\centering
\includegraphics[width=0.8\textwidth, height=\textheight, keepaspectratio]{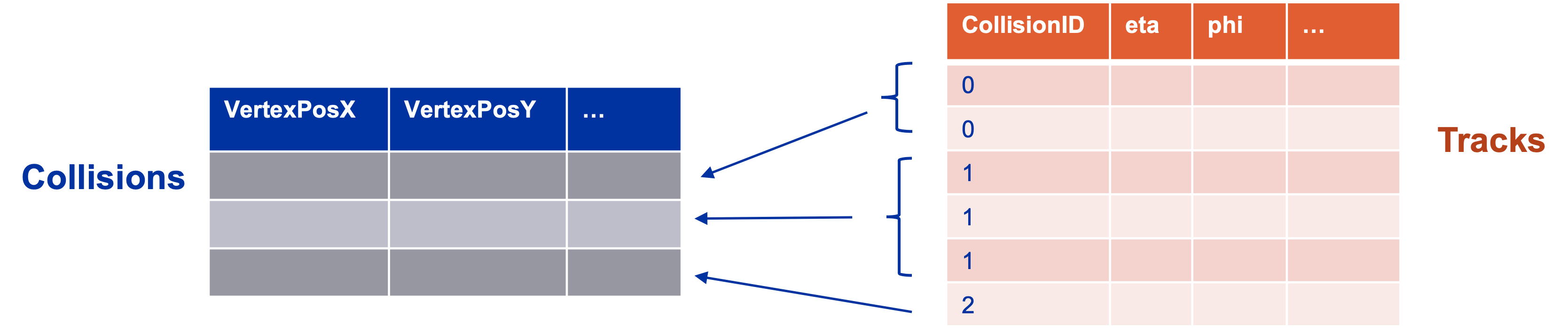}
  \caption{Interconnected tables}
  \label{fig:interonnectedtables}
\end{figure}

The data model format is similar to relational databases, while the complexity of representation is shielded from the user. For example, as illustrated in Figure~\ref{fig:interonnectedtables}, collisions and tracks are represented in trees (flat tables) that are connected through indices passed through shared memory.

A table is defined as a unique C++ type, templated on columns. The columnar format is provided by Apache Arrow~\cite{ApacheArrow}, which provides a contiguous in-memory layout for large amounts of data and ensures inter-process communication. This allows fast and efficient operations that, instead of being applied per collision or event, \mbox{they work per large data blocks with thousands of collisions.}

\begin{figure}[ht]
\includegraphics[width=\textwidth, height=\textheight, keepaspectratio]{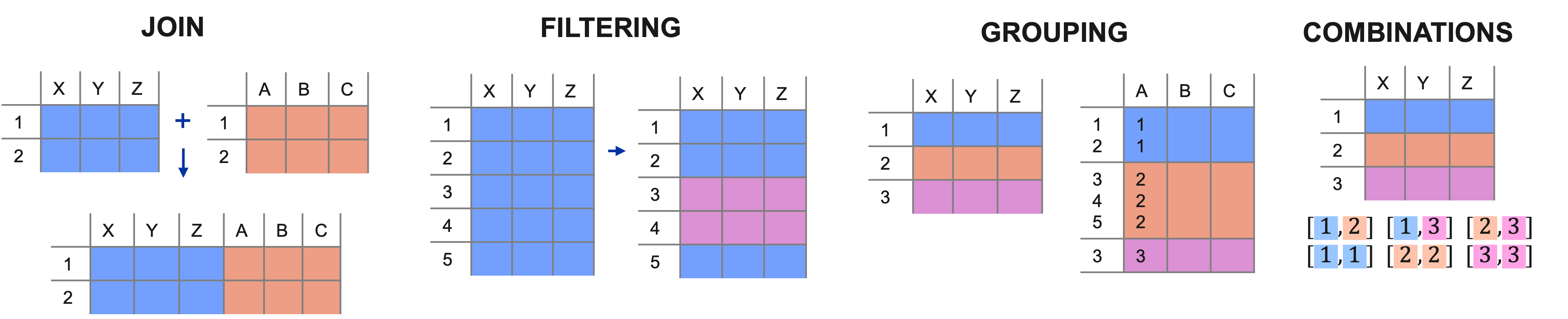}
  \caption{Table operations}
  \label{fig:operations}
\end{figure}

Joining, grouping, partitioning and filtering are requested by the analyzers and the operations can be combined when needed (Figure~\ref{fig:operations}). Apache Arrow ensures zero-copy operations, which means that the underlying contiguous arrays in memory are immutable, therefore no data is removed or copied in common operations.

\subsection{Data Processing}

\begin{figure}[ht]
\centering
\includegraphics[width=0.7\textwidth, height=\textheight, keepaspectratio]{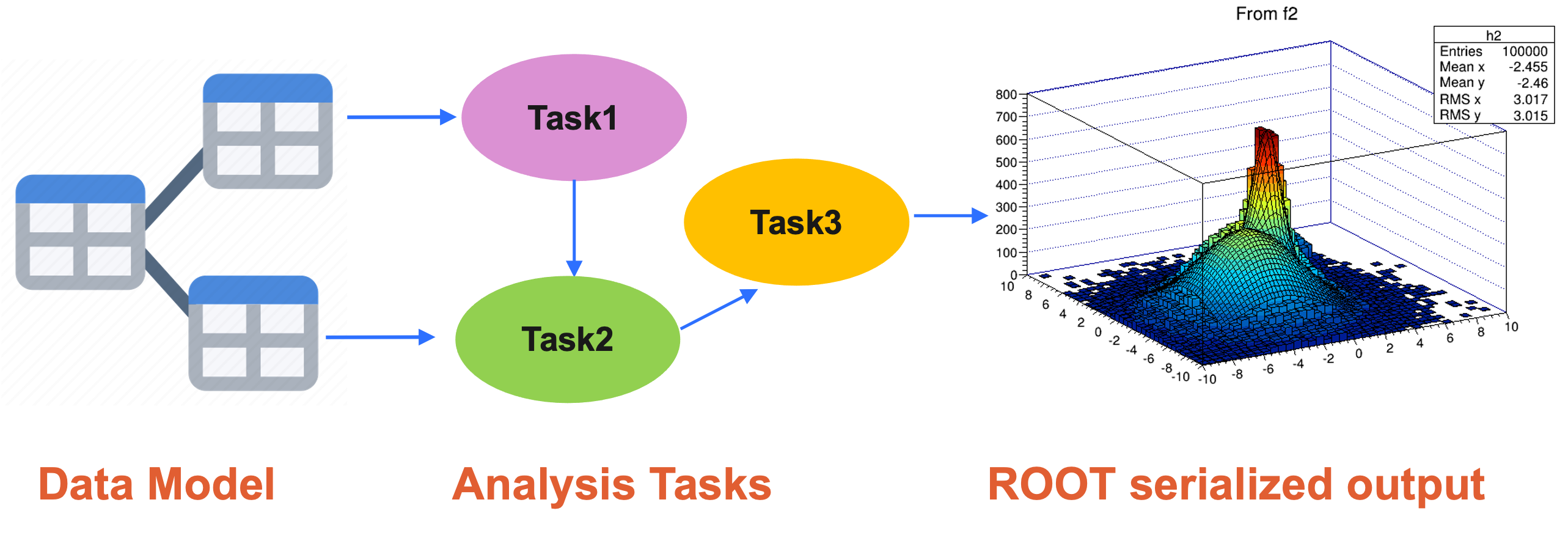}
  \caption{Data processing}
  \label{fig:dataprocessing}
\end{figure}

The {$\mathrm{O^2}$} framework ensures multiprocess capability. The user defines tasks that include callbacks and declarations of input and outputs, thus subscribing to different tables. These tasks are then combined into workflows representing a particular analysis or group of analyses, and the final result is the ROOT serialized output. This is illustrated in Figure~\ref{fig:dataprocessing}.

The analysis model is both declarative and imperative. The declarative way provides clear and compact definition of analysis cuts and flow, while the imperative way is used to represent certain conditions that cannot be implemented in a declarative way.
 
The Data Processing Layer hides the complexity of the abstract transport layer and low-level data model, builds the workflow topology based on interdependencies between tasks and translates the defined workflows to an actual FairMQ topology of devices.

\section{Hyperloop Train System}
\begin{figure}[ht]
\centering
\includegraphics[width=0.75\textwidth, height=\textheight, keepaspectratio]{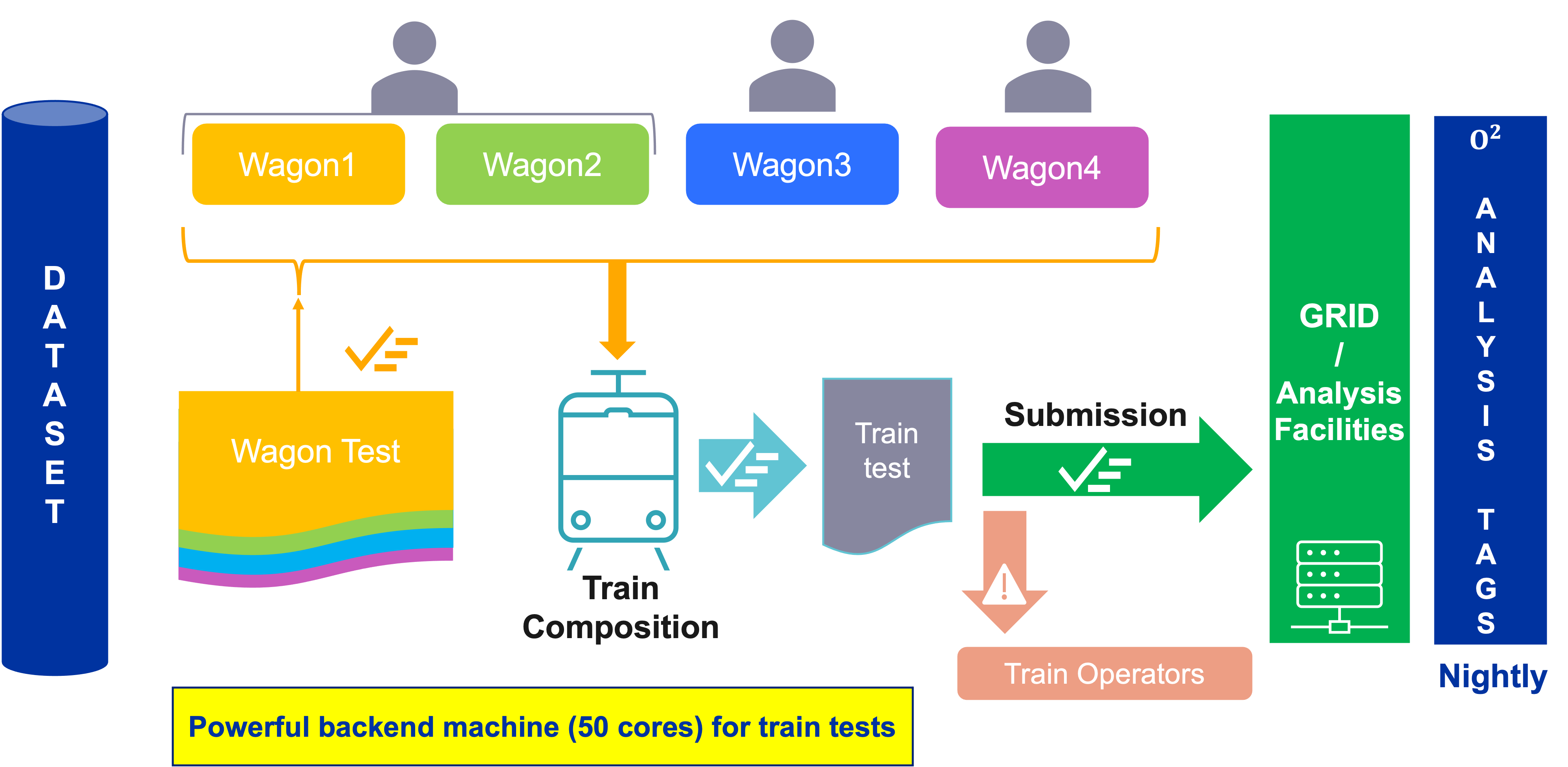}
  \caption{Analysis process in Hyperloop}
  \label{fig:hyperloopsummary}
\end{figure}

Hyperloop implements the concept of analysis trains used to optimize the usage of computing resources and was built upon tools used in Runs 1 and 2. It allows task configuration, enables fast and demanding analysis workflows on Grid and Analysis Facilities, which are specialized Grid sites with CPU and disk resources adjusted for analysis needs. A sketch of the analysis process in Hyperloop is shown in Figure~\ref{fig:hyperloopsummary}.

In Run 2, 90\% of the analysis activity used the so-called LEGO analysis trains~\cite{Zimmermann:2015owa}, which utilized on average about 40 000 cores at any given time. This tool was re-written with a modern reactive front-end technology, React~\cite{React}, and fully integrated with {$\mathrm{O^2}$}, providing efficiency, user-friendly features, and dedicated views for users and operators.

\subsection{Wagon tests and analysis process}

\begin{figure}[ht]
\centering
\includegraphics[width=0.95\textwidth, height=\textheight, keepaspectratio]{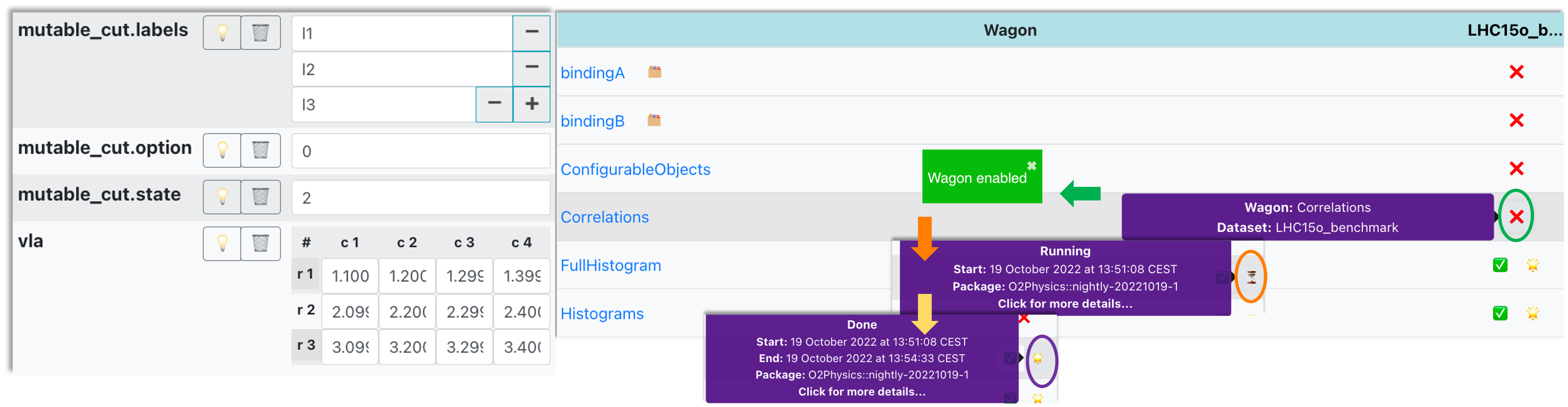}
  \caption{Wagon configuration (Left), Wagon test process (Right)}
  \label{fig:wagonConf}
\end{figure}

The data available in Hyperloop is converted Run 2 data, Run 3 data and MC and derived data. The user enables datasets per analysis (an organizational unit shared by multiple users) and is then able to create individual workflows, known as wagons. As illustrated in Figure~\ref{fig:wagonConf}, the wagon configuration supports a variety of parameter types such as primitive types, arrays, matrices, labelled matrices, and histogram binning. Subwagons can also be added and will run the same task with different parameter values, for instance for the evaluation of systematic uncertainties. Furthermore, output can be stored as derived data which is then used by a subsequent train (so-called tuple creation or skimming).

Once a wagon is enabled, a functional test will start immediately and the resulting information will be available in the wagon test view. This displays per device (reader, workflows, writer) and total performance metrics, as well as expected resources in interactive graphs. If the wagon test is successful, it will be composed together with wagons of other users into a train (either automatically or manually by an operator). If the train run test is successful, it will be submitted to the Grid or to an analysis facility (Figure~\ref{fig:hyperloopsummary}).

\subsection{Train Runs}
\begin{figure}[ht]
\centering
\includegraphics[width=0.7\textwidth, height=\textheight, keepaspectratio]{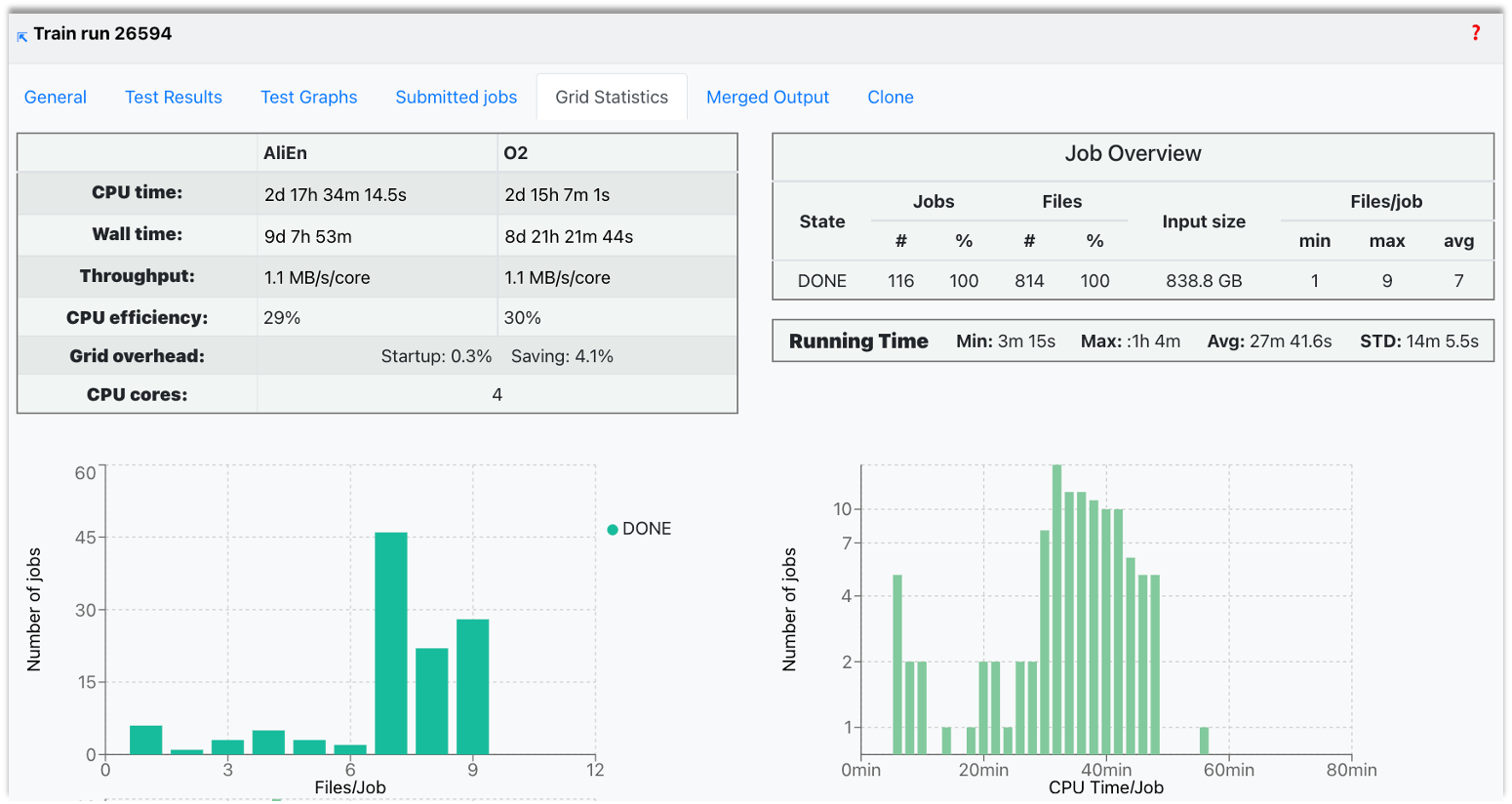}
  \caption{Grid statistics}
  \label{fig:gridStats}
\end{figure}

An automatic train composition schedule can be defined per dataset, and, if enabled, will compose trains at specified times based on target memory, wagon configuration and dependencies. After the train test is successful, it will be automatically submitted. The user can see the information about their train runs in a dedicated train result view. This displays the test results, the summary of submitted jobs and several statistics in interactive graphs (Figure~\ref{fig:gridStats}).

If automatic composition is not enabled for a dataset, users can request train composition on the dedicated operation channel. This request will be taken care of by one of the operators who work on a 24/5 shift-type basis (provided by four ALICE institutes spread across time zones). Besides helping with train composition, the operators offer support for warnings, failed tests and trains which are stuck due to specific problems. There is a dedicated page for trains which show pathological behaviour, which are then followed up by the operators. This allows a timely escalation to, for example, Grid experts if needed.

\subsection{Bookkeeping and monitoring}
History views are available for wagons and datasets, which allow to track all changes made. The user is also able to compare train and wagon configurations. Updates regarding these or other crucial elements in Hyperloop are sent to the user through a system of automatic notifications. Each notification contains an explanation of the update and links to details and documentation. 

The current status of the system is displayed in a dedicated dashboard. This is used by experts to track information such as number of wagons waiting to be tested and trains ready to be submitted. Moreover, it presents the summary of the previous week, showing the jobs' number and state per~site.

\section{Summary}
The ALICE detector received several major upgrades, allowing for 100 times more data to be recorded in Run 3 and 4. Considering the challenge to handle this great amount of data, new tools were developed: the {$\mathrm{O^2}$} Framework and the Hyperloop train system. The {$\mathrm{O^2}$} framework allows for distributed and efficient processing of the unprecedented data. It will hide the underlying complexity from the users, allowing them to focus on the physics analysis.

The Hyperloop train system enables analysis workflows to be run on the Grid and Analysis Facilities. Comparing to previous tools, it involves significant levels of automatic activity (e.g. train composition and submission), which translates into less manual actions to be taken by the operators. The modern interactive user interface offers a responsive, reliable and user-friendly application, and offers easy access to status and updates.

User support is carried out by operators that work on a 24/5 shift basis schedule. A dedicated channel is used for requests, questions and general updates. Moreover, extensive documentation has been created, which is easily accessible from several sections in Hyperloop and explains in detail the components, functions and processes that are part of the new tools. 

Hyperloop is in production and presently being used by 160 users in about 130 analyses. 133 datasets are available for processing and 4970 train runs have been submitted. Taking everything into consideration, the ALICE analysis infrastructure is ready for Run 3.

\end{document}